\documentclass[runningheads]{llncs}

\usepackage[T1]{fontenc}
\usepackage{amsmath,amssymb,amsfonts}
\usepackage{graphicx}
\usepackage{booktabs}
\usepackage{array}
\usepackage{enumitem}
\usepackage{url}
\usepackage{xcolor}
\usepackage{tikz}
\usetikzlibrary{arrows.meta,positioning,shapes.geometric,calc,fit}

\setlist[itemize]{leftmargin=*,topsep=2pt,itemsep=1pt}
\setlist[enumerate]{leftmargin=*,topsep=2pt,itemsep=1pt}

\definecolor{pyblue}{RGB}{0,90,160}
\definecolor{pylight}{RGB}{235,244,252}
\definecolor{pygray}{RGB}{80,80,80}
\definecolor{pylightgray}{RGB}{247,247,248}

\title{A Certified Higher-Order Quantum Framework for CSA and Margin-Aware Collateral Optimization}
\titlerunning{Certified Higher-Order Quantum Candidate Generation}

\author{Tao Jin\inst{1} \and Stuart Florescu\inst{2}}
\authorrunning{T. Jin and S. Florescu}
\institute{Pyligent AI\\
\email{tao.jin@pyligentai.com}
\and
Dept of Computing and Mathematical Sciences, California Institute of Technology\\
\email{sfloresc@caltech.edu}}

\begin{document}
\maketitle

\begin{abstract}
Collateral allocation for uncleared derivatives is a legally constrained, operationally discrete, and economically material optimization problem. Institutions must satisfy margin requirements while respecting Credit Support Annex (CSA) eligibility rules, valuation percentages, rounding, transfer thresholds, concentration limits, inventory availability, and variation-margin (VM), initial-margin (IM), or independent-amount (IA) side constraints. This manuscript develops \emph{CR-HO-QAOA}, a higher-order quantum candidate-generation framework for margin- and CSA-aware collateral allocation.

The framework is adapter-first. Official SIMM, proxy SIMM, legacy IA, VM-only, RQV, or hybrid margin sources are normalized upstream into a common margin-requirement object; the optimizer does not calculate or replace official SIMM. Given the normalized requirement, CSA-derived terms, and inventory, the optimization layer builds a bounded active neighborhood and maps its higher-order interactions to a Pauli-$Z$ cost Hamiltonian. The confirmatory neighborhoods are intentionally classically enumerable, so exact enumeration supplies local ground truth and native classical HOBO local search provides a necessary baseline. Sampled states are evaluated through a frozen deterministic certification table whose implemented decisions are cross-validated against fixed-allocation CP-SAT.

The archived v6 evaluation uses eleven primary fixtures with $N_{\mathrm{eff}}=8$, thirty seeds, a candidate budget of 16, and 6{,}300 seed-level rows under a preregistered gate protocol. Original builder idle-XY CR-HO-QAOA does \emph{not} outperform matched X-mixer QAOA on the primary certified metric: idle-XY is favored on 2/11 fixtures, X on 9/11, with mean idle-minus-X effect approximately $-0.04857$. At budget 16, mean improving-sample rates are $12.12\%$ for idle-XY and $16.85\%$ for matched X; native HOBO local search remains stronger on the registered classical comparison. A v6.2 amendment cross-validates the NumPy reduced-sector simulator against PennyLane \texttt{default.qubit} on all eleven fixtures at fixed $p=1$ parameters, with minimum fidelity $0.9999999999999991$. The study supports a reproducible formulation, simulator implementation, and governed certification architecture; it does not establish quantum advantage, runtime advantage, full collateral-constraint preservation, or production savings.
\end{abstract}

\section{Introduction}
\label{sec:introduction}

Collateral optimization is a natural candidate problem for quantum optimization because it is discrete, constrained, multi-objective, and economically material. In uncleared derivatives, a financial institution must select collateral assets that satisfy a credit support annex (CSA), variation margin (VM), initial margin (IM), eligibility schedules, haircut rules, concentration limits, inventory limits, custody restrictions, and operational movement costs. In SIMM-aware IM workflows, a regulatory margin engine first converts CRIF risk sensitivities into an IM requirement; the collateral optimizer must then satisfy that requirement using eligible and often segregated collateral.

The production bank process is therefore layered. Trades and risk systems produce sensitivities, usually exchanged through the Common Risk Interchange Format (CRIF); a SIMM calculator converts those sensitivities into an IM requirement; CSA or IM CSA terms define the legal collateral rules; the collateral inventory defines what can be posted; and the optimization layer chooses a certified allocation. This separation is central to the framework. The inventory list is not an input to SIMM. The CSA does not itself compute SIMM. The optimizer consumes the resulting required margin amount and legal collateral rules, then solves the downstream collateral allocation problem.

The quantum-finance literature has already recognized collateral optimization as a combinatorial optimization problem. Giron et al. formulate collateral allocation as both a mixed-integer linear program and a QUBO, explicitly positioning it for NISQ and quantum-inspired computation~\cite{giron2023collateral}. Our prior Pyligent V1 study reported exploratory improvements for a certified hybrid pipeline combining evidence-gated CSA extraction, simulated annealing, micro-HO-QAOA on binding sub-QUBOs, and CP-SAT certification~\cite{jin2025hybrid}. Those exploratory results motivated, but do not provide confirmatory evidence for, the present quantum-algorithmic study. The question addressed here is how to replace an implementation-oriented sub-QUBO hook with an explicit higher-order Hamiltonian, structured-subspace mixer design, circuit-level analysis, and frozen classical controls.

A plain QUBO is not always the natural model for real collateral workflows. Production decisions contain higher-order interactions: three securities can jointly breach a currency cap; two recall/post actions can activate the same triparty instruction; a substitution can be cheap in funding terms but costly after IM segregation; and lot-size or CSA-rounding effects can make small surplus acceptable while shortfall remains prohibited. Reducing all such structure to pairwise penalties can increase auxiliary-variable count, distort penalty calibration, and reduce the probability of sampling useful feasible candidates.

We propose a quantum optimization framework that treats collateral allocation as a \emph{higher-order local search problem} rather than as a monolithic global QUBO. The proposed method, \emph{CR-HO-QAOA} (a certified higher-order / structured-subspace QAOA formulation), starts from a feasible allocation produced by a deterministic master solver, extracts a compact active neighborhood of collateral actions, constructs a higher-order binary optimization (HOBO/HUBO) Hamiltonian, and applies a collateral-informed QAOA ansatz to sample local improvements. Confirmatory neighborhoods are intentionally classically enumerable so that exact enumeration supplies local ground truth; we do not claim a quantum computational advantage at this scale. In the production architecture, sampled candidates are decoded, optionally repaired, and submitted to deterministic certification before any recommendation. In the confirmatory benchmark, every local action state is pre-evaluated by the implemented deterministic checker, and those decisions are cross-validated against fixed-allocation CP-SAT. To our knowledge, CR-HO-QAOA is among the first higher-order, structured-subspace QAOA formulations for CSA- and margin-requirement-aware collateral allocation with an adapter-first SIMM/legacy-margin boundary and an explicit certification audit.

The novelty is therefore not that a quantum circuit replaces the margin engine or the deterministic optimizer. The novelty is a domain-specific quantum optimization layer with four design choices:
\begin{enumerate}
    \item \textbf{Higher-order cost Hamiltonian.} Collateral actions are represented as a hypergraph whose edges encode funding, movement, coverage, concentration, segregation, and governance interactions of order up to four.
    \item \textbf{Structured-subspace mixers.} Instead of relying only on a transverse-field mixer and large penalties, the framework proposes XY-style and substitution-aware mixers for subspaces such as one-hot alternatives, fixed-cardinality movement budgets, VM/IM side assignment, and recall-post pair structure. The confirmatory experiment distinguishes these proposed families from the executed builder idle-XY mixer and reports that full joint constraint preservation fails (Gate~B2).
    \item \textbf{Warm-started local quantum search.} The initial state is concentrated around a CP-SAT feasible allocation or current production allocation, reflecting the operational reality that collateral desks usually seek controlled reallocations, not unconstrained global reshuffling.
    \item \textbf{Certified hybrid execution.} The quantum layer is used for candidate generation. The production design retains CP-SAT as the deterministic system of record; confirmatory scoring uses a frozen deterministic certification table cross-validated against fixed-allocation CP-SAT.
\end{enumerate}

This positioning is aligned with recent quantum optimization research. Higher-order portfolio optimization shows that realistic financial objectives can naturally generate HUBO terms rather than QUBO terms~\cite{uotila2025higher}. Quantum approximate multi-objective optimization is relevant because collateral allocation is a tradeoff among funding, movement, liquidity, concentration, and governance objectives~\cite{kotil2025qamoo}. Constraint-preserving QAOA with XY mixers is relevant because hard financial constraints are poorly served by penalty-only formulations~\cite{hadfield2019qaoa,kordonowy2026xy,mancilla2026constrained}. Warm-starting and hot-starting methods are relevant because financial solvers often begin from a high-quality incumbent rather than from an unstructured search distribution~\cite{egger2021warm,schlutter2025hot}.

Our work makes the following contributions:
\begin{itemize}
    \item a quantum-algorithmic extension of the Pyligent V1 CSA optimization pipeline~\cite{jin2025hybrid}, moving from a certified hybrid sub-QUBO prototype to an explicit higher-order QAOA model;
    \item an adapter-first boundary between upstream margin calculation and a normalized downstream requirement;
    \item a higher-order Pauli Hamiltonian formulation for local collateral reallocation;
    \item a one-sided coverage slack ladder that distinguishes unacceptable margin shortfall from operationally acceptable surplus;
    \item collateral-informed mixer families, with an explicit confirmatory distinction between proposed families and the executed builder idle-XY mixer;
    \item a warm-started HO-QAOA workflow for active-neighborhood optimization under near-term quantum limits;
    \item a reproducible confirmatory evaluation with exact enumeration, matched X-mixer QAOA, and native HOBO local search under a preregistered gate protocol;
    \item exact local-state pre-evaluation, an implemented-constraint audit, and fixed-allocation CP-SAT cross-checks;
    \item a negative empirical finding: the original builder loses to matched X-mixer QAOA on the primary metric, while native HOBO local search remains stronger.
\end{itemize}

\subsection{Positioning and Innovation in the Collateral Optimization}
\label{subsec:positioning_2026}

The contribution of this work is not to replace existing bank risk engines, or collateral workflow platforms. In production, ISDA SIMM remains the appropriate upstream methodology for calculating regulatory initial margin, and banks may continue to rely on approved internal or vendor SIMM engines. The proposed framework instead begins after margin-source normalization: official SIMM outputs, CRIF-driven proxy margins, or legacy margin requirements are converted into a common \texttt{MarginRequirement} object. The collateral optimizer then consumes only the normalized required amount, together with CSA-derived legal terms and inventory data.

This distinction is important for bank deployment. A real collateral decision is not determined by the SIMM number alone. The SIMM or legacy margin requirement defines how much collateral must be posted, while the CSA defines what collateral is legally eligible, how it is valued, what haircuts apply, what rounding and minimum transfer rules govern delivery, and whether segregation or rehypothecation restrictions apply. The inventory system then defines what collateral is actually available. Pyligent therefore targets the decision layer that connects these inputs: legal agreement terms, margin requirements, and available collateral inventory.

The main innovation is a CSA- and margin-source-aware collateral optimization framework that treats real collateral allocation as a higher-order constrained decision problem. Unlike generic portfolio optimization, collateral management contains discontinuities and bundle effects that are not naturally pairwise. Examples include settlement-batch costs, concentration limits, current-allocation substitutions, chunky collateral lots, overshoot from rounding, minimum transfer amounts, and IM segregation or non-reuse costs. These interactions motivate a higher-order binary optimization formulation rather than a purely quadratic QUBO representation.

The proposed CR-HO-QAOA layer is designed as a structured candidate-generation mechanism for these higher-order collateral interactions. It is not treated as an autonomous production decision-maker. Quantum-generated candidates are decoded, repaired if necessary, evaluated under the production objective, and subjected to deterministic certification before any recommendation is reported. In production, CP-SAT is the intended final arbiter. In the confirmatory package, samples are looked up in a frozen table produced by the deterministic checker, with fixed-allocation CP-SAT used to validate equivalence for the implemented constraint set. Confirmatory evidence does not support an optimization advantage for the original builder over matched X-mixer QAOA or native HOBO local search.

This architecture provides three practical advantages. First, it is compatible with bank and vendor SIMM engines because the optimizer does not need to implement official SIMM internally. Second, it supports both modern SIMM/CRIF workflows and legacy pre-SIMM collateral processes through the same normalized margin interface. Third, it produces audit-ready outputs: margin-source provenance, CSA evidence, inventory hashes, objective breakdowns, higher-order Hamiltonian metadata, deterministic certification records, and CP-SAT cross-checks. As a result, the framework positions quantum optimization as part of a governed collateral decision pipeline rather than as a black-box replacement for existing bank systems.

In summary, the novelty of this work lies in combining adapter-first margin-source normalization, CSA-aware higher-order collateral modeling, structured-subspace quantum candidate generation, and deterministic certification with CP-SAT validation. This combination is intended to make quantum optimization relevant to a concrete financial-infrastructure problem while keeping legal feasibility and production authority outside the variational sampler.

Table~\ref{tab:old_new} summarizes the central algorithmic delta relative to the earlier QUBO-style local-search hook. It is placed here because the manuscript's contribution is not merely a finance use case; it is the replacement of a quadratic penalty sampler with a higher-order, structured-subspace quantum local-search design whose confirmatory value is architectural and formulational rather than an optimization-superiority claim.

\begin{table}[t]
\centering
\caption{Algorithmic design delta from old QUBO-style local search to CR-HO-QAOA (formulation intent; not a confirmatory performance claim).}
\label{tab:old_new}
\small
\begin{tabular}{>{\raggedright\arraybackslash}p{0.24\linewidth}>{\raggedright\arraybackslash}p{0.27\linewidth}>{\raggedright\arraybackslash}p{0.37\linewidth}}
\toprule
\textbf{Design issue} & \textbf{Old QUBO-style approach} & \textbf{CR-HO-QAOA approach} \\
\midrule
Model structure & Quadratic penalty model & Higher-order collateral hypergraph \\
Quantum cost & Pairwise Ising terms & Pauli-$Z$ strings from HOBO hyperedges \\
Coverage & Symmetric surplus/shortfall penalty & One-sided slack ladder \\
Mixer & Transverse-field or sampler-only & Proposed side, one-hot, cardinality, substitution, and batch families; frozen run uses idle-XY and matched X \\
Initialization & Random or uniform & CP-SAT or current-allocation warm start \\
Search region & Broad binary space & Binding active neighborhood, $N\le16$ \\
SIMM treatment & Generic requirement input & Adapter-first \texttt{MarginRequirement}; IM-side cost, segregation, and attribution-aware neighborhood selection \\
Production safety & Post-hoc checking & Deterministic output gate; CP-SAT is production authority \\
\bottomrule
\end{tabular}
\end{table}

\section{Related Work}
\label{sec:related}

\subsection{Collateral Optimization and QUBO Baselines}
Collateral allocation is classically modeled using integer or mixed-integer programming. The most directly related quantum-finance work is Giron et al.~\cite{giron2023collateral}, which develops a MILP formulation and then translates the collateral problem into a QUBO suitable for NISQ and quantum-inspired experimentation. This paper builds on that direction but focuses on the gap between QUBO demonstration and production collateral structure. The proposed algorithm builds a local higher-order Hamiltonian and uses structured-subspace quantum dynamics to generate candidate distributions. The confirmatory results do not demonstrate improved feasibility, diversity, or optimization quality over the matched controls.

The immediate predecessor to this work is the Pyligent V1 paper on hybrid LLM and HO-QAOA for CSA collateral management~\cite{jin2025hybrid}. V1 demonstrated an exploratory end-to-end pipeline: CSA clauses were extracted into governed JSON with span evidence; collateral decisions were optimized using simulated annealing interleaved with micro-HO-QAOA on binding subproblems; and CP-SAT was used as a deterministic arbiter. Those results motivate, but are not evidence for, the present confirmatory claims. Here the quantum component is the central object of study: the local subproblem is formulated as a collateral hypergraph, mapped to a Pauli-$Z$ Hamiltonian, evolved with an executed builder idle-XY ansatz, and compared with matched X-mixer QAOA and native HOBO local search on frozen $N_{\mathrm{eff}}=8$ fixtures.

\subsection{QAOA, Alternating Operators, and Feasible Subspaces}
QAOA was introduced as a variational algorithm for approximate combinatorial optimization~\cite{farhi2014qaoa}. The more general quantum alternating operator ansatz extends QAOA by allowing custom phase separators and mixers matched to the problem domain~\cite{hadfield2019qaoa}. This generalization is essential for finance applications where hard constraints cannot be treated as merely soft preferences.

A standard transverse-field mixer,
\begin{equation}
H_X=\sum_i X_i,
\end{equation}
explores the entire Boolean hypercube and therefore frequently leaves the feasible region. In collateral optimization, infeasible states can violate eligibility, inventory, segregation, or cardinality requirements. Penalty terms can suppress such states, but large penalties distort the energy landscape and can make optimization unstable. Constraint-preserving mixers solve this problem by restricting evolution to a feasible subspace. Recent XY-mixer and constrained-portfolio studies motivate this design for financial selection problems with cardinality or budget constraints~\cite{kordonowy2026xy,mancilla2026constrained}.

\subsection{Higher-Order Binary Optimization in Quantum Finance}
Many quantum finance examples use QUBO because pairwise Ising Hamiltonians are simple to implement. However, realistic finance can naturally produce higher-order interactions. Uotila et al.~\cite{uotila2025higher} show that portfolio optimization with skewness and kurtosis leads to higher-order terms and a HUBO formulation. Collateral optimization exhibits an analogous phenomenon, not through higher statistical moments, but through legal and operational interactions among groups of collateral actions. A batch instruction, issuer cap, currency cap, or IM segregation threshold can depend on a combination of decisions rather than on any single security.

\subsection{SIMM, CRIF, and the Governance Boundary}
ISDA SIMM is the standard model for calculating regulatory IM for non-cleared derivatives, and CRIF is the standard sensitivity exchange format used in SIMM workflows~\cite{isda_simm_infohub,isda_crif,isda_simm_methodology_282506,isda2025simm,isda_simm_trusted}. The CSA side of the legal boundary is anchored in ISDA's 2016 VM Credit Support Annex documentation and the 2016 Variation Margin Protocol~\cite{isda_2016_vm_csa_ny,isda_2016_vm_csa_english,isda_2016_vm_protocol}. Regulatory validation of SIMM increases the need to keep the margin engine separate from the collateral optimizer~\cite{eba2026simm}.

This paper adopts an adapter-first SIMM boundary. In production, a bank may provide CRIF plus an official SIMM result from an internal engine, a vendor engine, or another approved platform. In research and software validation, a CRIF-driven demo proxy may produce a non-official IM requirement, but such results must be labeled as proxy outputs. In all cases, the result is normalized into a margin requirement object before optimization. CRIF sensitivities are not quantum decision variables, and inventory assets are not inputs to SIMM. The quantum optimizer solves the downstream collateral allocation problem conditional on the normalized requirement, CSA-derived collateral rules, and inventory snapshot.

The quantum-application gap addressed by this paper is therefore precise: prior collateral QUBO work shows feasibility of quantum-ready encoding; CR-HO-QAOA proposes a higher-order, structured-subspace, warm-started quantum candidate-generation layer suitable for CSA/SIMM and legacy margin workflows without claiming to replace the approved margin engine or deterministic certification authority.

\section{Financial Problem Setting: Collateral as the Domain Generator}
\label{sec:domain}

The financial domain defines the legal and economic constraints that generate the quantum optimization instance. In this paper, VM terms are aligned with the 2016 ISDA VM CSA documentation, while SIMM-related metadata and CRIF processing are aligned with ISDA SIMM and CRIF standards~\cite{isda_2016_vm_csa_ny,isda_2016_vm_csa_english,isda_simm_methodology_282506,isda_crif}. The optimizer, however, is deliberately margin-source agnostic.

\subsection{Bank Process and Adapter-First Margin Boundary}
In a bank deployment, collateral optimization occurs after margin calculation. Trades are valued by front-office and risk systems; sensitivities are exported, commonly as CRIF; a SIMM engine or equivalent margin service calculates IM; CSA or IM CSA terms define the legal collateral rules; and the inventory system provides the available collateral pool. Pyligent's optimization layer begins after the required amount is known. It may consume an official external SIMM result, a bank internal SIMM result, a vendor result, a clearly labeled demo proxy, or a legacy margin calculation. It then normalizes the result into a \texttt{MarginRequirement}.

The normalized requirement is
\begin{equation}
\mathcal{R}_n=(R_n^{VM},R_n^{IM},R_n^{IA},R_n,M_n,\nu_n,\Xi_n),
\end{equation}
where $R_n^{VM}$ is the VM requirement, $R_n^{IM}$ is the SIMM or IM requirement, $R_n^{IA}$ is any legacy independent amount, $R_n$ is the total amount to be covered, $M_n$ is the margin route, $\nu_n$ is the SIMM or legacy version label when applicable, and $\Xi_n$ contains source metadata such as CRIF hash, CSA hash, valuation date, calculation source, and official/proxy flags. The total required amount is
\begin{equation}
R_n = R_n^{VM}+R_n^{IM}+R_n^{IA},
\end{equation}
up to contract-specific hybrid rules such as \texttt{sum}, \texttt{max}, SIMM-only, or legacy-only treatment.

\subsection{Legal and Collateral State}
Let $n$ denote a netting set. The CSA and margin preprocessing layer produces a normalized state
\begin{equation}
\mathcal{L}_n=(M_n,\Pi_n,\Theta_n,\mathcal{C}_n,\mathcal{H}_n,\mathcal{G}_n),
\end{equation}
where $M_n$ is the legal margin-method route, $\Pi_n$ is a product-level SIMM/Grid routing map, $\Theta_n$ contains thresholds, MTA, rounding, and transfer mechanics, $\mathcal{C}_n$ is the side-specific eligibility set, $\mathcal{H}_n$ is the haircut rule set, and $\mathcal{G}_n$ stores governance metadata.

The route is
\begin{equation}
M_n\in\{\mathrm{VM},\mathrm{SIMM},\mathrm{Grid},\mathrm{LegacyIA},\mathrm{RQV},\mathrm{Hybrid},\mathrm{Abstain}\}.
\end{equation}
If the route is \texttt{Abstain}, no optimization is run. If the route is executable, the margin layer returns the normalized requirement $\mathcal{R}_n$. This abstraction supports SIMM/CRIF, external official SIMM, legacy independent amount grids, VM-only agreements, old RQV or exposure workflows, and hybrid cases. The downstream objective, HOBO construction, QAOA ansatz, and CP-SAT certifier use $R_n$ and the route metadata, but they do not recompute SIMM.

For SIMM-routed cases, the IM amount is generated upstream:
\begin{equation}
IM_n^{gross}=\mathcal{M}_{SIMM}^{\nu_n}(\mathcal{S}_n),
\end{equation}
where $\nu_n$ is the SIMM methodology version and $\mathcal{S}_n$ is the controlled CRIF sensitivity set. The function $\mathcal{M}_{SIMM}^{\nu_n}$ may represent an external official bank or vendor engine in production, or a clearly labeled non-official demo proxy in experiments. The optimizer consumes the resulting requirement,
\begin{equation}
R_n^{IM}=\mathrm{Round}_{RA^{IM}}\left(\max(0,IM_n^{gross}-T_n^{IM})\right),
\end{equation}
where $T_n^{IM}$ is the IM threshold and $RA^{IM}$ is the IM rounding amount when applicable. The inventory list is not used to calculate $IM_n^{gross}$; it is used after this step to select eligible collateral.

Collateral block $i$ has unit value $p_i$, side-specific haircut $h_{i,s}$, availability $A_i$, and eligibility indicator $\eta_{i,s}\in\{0,1\}$ for side $s\in\{VM,IM\}$. Its haircut-adjusted unit value is
\begin{equation}
a_{i,s}=p_i(1-h_{i,s}).
\end{equation}
For a full allocation $q_{i,s}$, side coverage is
\begin{equation}
U_s(q)=\sum_i \eta_{i,s}a_{i,s}q_{i,s}.
\end{equation}
The hard coverage constraints are
\begin{equation}
U_{VM}(q)\ge R_n^{VM},\qquad U_{IM/IA}(q)\ge R_n^{IM}+R_n^{IA},
\end{equation}
with side assignment and segregation governed by the CSA, IM CSA, or legacy agreement. In VM-only cases, the second constraint is inactive; in hybrid cases, the route determines whether SIMM IM and legacy IA are summed, maximized, or selectively applied.

The production architecture assigns hard feasibility to a deterministic master problem rather than to the quantum sampler. Inventory, coverage, eligibility, side assignment, concentration, lot, custody, and segregation rules should be represented explicitly before deployment. The confirmatory implementation covers only the subset reported in Section~\ref{sec:benchmark}: some substitution, batch, custody, and segregation logic remains in the objective, decoding path, or metadata rather than in a hard CP-SAT predicate. The quantum algorithm receives a feasible baseline $q^0$ and searches a compact local neighborhood around it.

\section{Higher-Order Quantum Optimization Model}
\label{sec:quantum_model}

\subsection{Local Collateral Action Variables}
The quantum subproblem is constructed over local actions rather than over all inventory. Let
\begin{equation}
\mathcal{A}=\{1,\ldots,N\}
\end{equation}
be a set of candidate actions extracted from the CP-SAT baseline. An action may post a block, recall a block, substitute one asset for another, activate a custody batch, or select a surplus slack denomination. The binary variable
\begin{equation}
x_i\in\{0,1\}
\end{equation}
indicates whether action $i$ is selected.

Near-term quantum execution imposes a bounded active set. In this paper,
\begin{equation}
N \triangleq |\mathcal{A}|,
\qquad
K \triangleq \max_{e\in\mathcal{E}} |e|,
\end{equation}
where $N$ is the number of active local collateral-action variables embedded in the quantum subproblem and $K$ is the maximum hyperedge order. Importantly, $N$ is not the total number of collateral assets in the inventory. It is the size of the selected active neighborhood around the current or CP-SAT feasible allocation.

The proposed design and scaling grid is
\begin{equation}
N\in\{8,12,16\},\qquad K\in\{2,3,4\}.
\label{eq:nk_grid}
\end{equation}
These values define a prospective ablation grid, not the executed confirmatory scope and not a universal optimality theorem. The frozen confirmatory package reported in Section~\ref{sec:benchmark} uses $N_{\mathrm{requested}}=N_{\mathrm{eff}}=8$; diagnostic Hamiltonians include multiple registered orders, while no confirmatory $N=12$ or $N=16$ performance claim is made. The construction can be recursively applied across multiple active neighborhoods.

\subsection{$N$--$K$ Practicality and Non-Optimality Claim}
\label{subsec:nk_regime}

We do not claim that $N\le16$ and $K\le4$ are globally optimal across all collateral portfolios. A universal proof would require showing that no larger or differently structured neighborhood can ever improve any CSA/SIMM instance, which is not realistic because the best setting depends on inventory size, eligibility, haircuts, lot sizes, rounding rules, concentration caps, VM/IM split, SIMM IM requirement, funding curves, liquidity assumptions, and runtime limits. The weaker claim is that $N\in\{8,12,16\}$ and $K\in\{2,3,4\}$ form a bounded research grid suitable for exact local validation and analytical resource accounting. Only the $N_{\mathrm{eff}}=8$ confirmatory slice is executed in the frozen performance package.

For $N$ active binary variables, the raw search space has size
\begin{equation}
|\{0,1\}^N|=2^N.
\end{equation}
Thus $N=16$ gives $65{,}536$ local states, which is classically trivial to enumerate on contemporary hardware. Such sizes are useful here only because exact local ground truth permits verification of sampler probabilities, objective alignment, feasible-state mass, and certified candidate quality; they provide no evidence of computational advantage.

\begin{table}[t]
\centering
\caption{Search-space growth for the active quantum neighborhood.}
\label{tab:n_search_space}
\begin{tabular}{rr}
\toprule
\textbf{Active variables $N$} & \textbf{Raw states $2^N$} \\
\midrule
8  & 256 \\
12 & 4{,}096 \\
16 & 65{,}536 \\
20 & 1{,}048{,}576 \\
24 & 16{,}777{,}216 \\
\bottomrule
\end{tabular}
\end{table}

The number of possible hypergraph terms up to order $K$ is
\begin{equation}
T(N,K)=\sum_{r=1}^{K}\binom{N}{r}.
\label{eq:hyp_term_count}
\end{equation}
For $N=16$, this gives the growth shown in Table~\ref{tab:k_term_growth}. Moving from $K=2$ to $K=3$ and $K=4$ materially increases modeling expressiveness; moving beyond $K=4$ rapidly increases Hamiltonian and circuit complexity. Therefore $K\le4$ is a practical cutoff for a near-term higher-order QAOA study.

\begin{table}[t]
\centering
\caption{Maximum number of hypergraph terms for $N=16$.}
\label{tab:k_term_growth}
\begin{tabular}{rr}
\toprule
\textbf{Maximum order $K$} & \textbf{Terms $T(16,K)$} \\
\midrule
2 & 136 \\
3 & 696 \\
4 & 2{,}516 \\
5 & 6{,}884 \\
\bottomrule
\end{tabular}
\end{table}

The collateral interpretation of $K$ is also economically grounded. Linear terms capture individual asset economics; pairwise terms recover the standard QUBO baseline; third- and fourth-order terms capture group effects that are common in CSA collateral workflows.

\begin{table}[t]
\centering
\caption{Collateral interpretation of the interaction order $K$.}
\label{tab:k_interpretation}
\begin{tabular}{p{0.13\linewidth}p{0.73\linewidth}}
\toprule
\textbf{Order} & \textbf{CSA/SIMM collateral meaning} \\
\midrule
$K=1$ & Individual funding, haircut-adjusted value, liquidity, custody, or segregation cost. \\
$K=2$ & Pairwise QUBO effects, such as two assets jointly pressuring an issuer, currency, or custodian cap. \\
$K=3$ & Operational and legal group effects, such as two substitutions activating a common settlement batch or interacting with a surplus-slack denomination. \\
$K=4$ & Higher-order CSA/SIMM structure, such as concentration, currency, IM segregation, and liquidity bucket effects appearing together in one collateral package. \\
\bottomrule
\end{tabular}
\end{table}

For a future full-grid deployment study, an instance-dependent setting could be selected by a resource-adjusted criterion,
\begin{equation}
(N^{\star},K^{\star})\in
\arg\min_{(N,K)\in\mathcal{G}}
\left[J(q_{N,K}^{cert})+\lambda_{\tau}\tau_{N,K}+\lambda_DD_{N,K}^{circ}\right],
\label{eq:nk_selection}
\end{equation}
where $\mathcal{G}=\{8,12,16\}\times\{2,3,4\}$, $J(q_{N,K}^{cert})$ is the deterministically certified collateral objective, $\tau_{N,K}$ is runtime or sampling cost, $D_{N,K}^{circ}$ is circuit-resource cost, and $\lambda_{\tau},\lambda_D$ encode resource penalties. This selection rule is a proposed extension; it was not used to choose the frozen confirmatory configuration and must not be interpreted as a completed $N$--$K$ experiment.

A local action set is selected using binding-domain signals:
\begin{itemize}
    \item currently posted collateral with high funding or liquidity cost;
    \item near-cheapest eligible substitutes;
    \item assets close to issuer, rating, currency, or custodian caps;
    \item lot-size actions that influence CSA rounding or MTA residuals;
    \item IM-eligible assets with high segregation or non-reuse cost;
    \item collateral candidates linked to large SIMM IM requirement or attribution buckets.
\end{itemize}

The full allocation induced by action vector $x$ is
\begin{equation}
q(x)=q^0+\Delta q(x),
\end{equation}
where $\Delta q(x)$ is the sum of selected local action deltas.

\subsection{Higher-Order Collateral Hypergraph}
Define a collateral hypergraph
\begin{equation}
\mathcal{H}=(\mathcal{V},\mathcal{E}),
\end{equation}
where vertices are local binary actions and each hyperedge $e\in\mathcal{E}$ connects a subset of actions $e\subseteq\mathcal{V}$ with order $|e|\le K$. The higher-order binary objective is
\begin{equation}
C(x)=\sum_{e\in\mathcal{E}} c_e\prod_{i\in e}x_i.
\label{eq:hubo_cost}
\end{equation}
This is a higher-order unconstrained binary optimization model. In the frozen implementation, the coefficients $c_e$ are calibrated from the eight-component search objective
\begin{equation}
C(x)=C_F(x)+C_M(x)+R_L(x)+P_O(x)+P_I(x)+P_C(x)+P_S(x)+P_G(x),
\label{eq:eight_component_objective}
\end{equation}
where $C_F$ is funding cost; $C_M$ covers movement, ticket, settlement, and substitution operational costs; $R_L$ is liquidity or tail-risk cost; $P_O$ is overshoot cost; $P_I$ is IM segregation and non-reuse cost; $P_C$ is concentration pressure; $P_S$ is a shortfall search and diagnostic penalty; and $P_G$ is governance or manual-review cost. The search penalty $P_S$ is not economically tradeable at certification: a prohibited positive shortfall rejects the candidate. The certified production objective is evaluated only on candidates that pass the implemented hard checks.

Table~\ref{tab:hyperedge_examples} gives examples of higher-order collateral hyperedges.

\begin{table}[t]
\centering
\caption{Higher-order collateral hyperedges and quantum optimization meaning.}
\label{tab:hyperedge_examples}
\begin{tabular}{p{0.26\linewidth}p{0.62\linewidth}}
\toprule
\textbf{Term} & \textbf{Meaning in collateral allocation} \\
\midrule
$x_i x_j x_k$ & Three selected collateral blocks jointly exceed an issuer, rating, currency, or custodian pressure threshold. \\
$x_i x_j b_m$ & Two movements activate the same custody, settlement, or triparty batch instruction $b_m$. \\
$x_i x_j s_{d}$ & A substitution pair interacts with a surplus slack denomination used to absorb CSA rounding excess. \\
$x_i x_j x_k x_{\ell}$ & A four-action combination crosses an IM segregation, liquidity-stress, or internal concentration threshold. \\
$\prod_{i\in G}(1-x_i)$ & None of the acceptable alternatives in group $G$ is selected. \\
\bottomrule
\end{tabular}
\end{table}

The higher-order representation is preferable when the business logic is inherently group-based. Quadratizing every interaction into auxiliary variables can be useful for QUBO backends, but it may obscure the financial semantics and enlarge the binary space. CR-HO-QAOA keeps the hypergraph representation as the primary mathematical object.

\subsection{Pauli-$Z$ Hamiltonian Mapping}
Each binary variable is mapped to a qubit by
\begin{equation}
x_i=\frac{1-Z_i}{2},
\end{equation}
where $Z_i$ is the Pauli-$Z$ operator on qubit $i$. A hyperedge term becomes
\begin{equation}
c_e\prod_{i\in e}x_i
=
\frac{c_e}{2^{|e|}}\prod_{i\in e}(I-Z_i)
=
\frac{c_e}{2^{|e|}}\sum_{S\subseteq e}(-1)^{|S|}\prod_{j\in S}Z_j.
\label{eq:pauli_expansion}
\end{equation}
The quantum cost Hamiltonian is therefore
\begin{equation}
\hat{H}_C=\sum_{e\in\mathcal{E}}\frac{c_e}{2^{|e|}}\sum_{S\subseteq e}(-1)^{|S|}\prod_{j\in S}Z_j.
\label{eq:cost_hamiltonian}
\end{equation}
For $K\le4$, each hyperedge expands into at most $16$ Pauli-$Z$ strings. This bound keeps the local phase separator executable on simulators and small hardware experiments.

\subsection{One-Sided Margin Coverage Slack}
A symmetric penalty $(U_s(x)-R_s)^2$ incorrectly treats collateral surplus and shortfall as equally undesirable. In collateral management, shortfall is unacceptable, while small surplus may be unavoidable because of lot size, CSA rounding, and eligible-inventory discreteness. We therefore encode coverage using nonnegative slack bits.

Let local side coverage be
\begin{equation}
U_s(x)=U_s(q^0)+\sum_i \Delta u_{i,s}x_i,
\end{equation}
where $s\in\{VM,IM\}$. Introduce binary surplus slack variables $z_{s,d}$ with denominations $\delta_d>0$:
\begin{equation}
S_s(z)=\sum_d\delta_dz_{s,d}.
\end{equation}
The side-specific coverage penalty is
\begin{equation}
P_{cov,s}(x,z)=\lambda_s\left(R_s-U_s(x)+S_s(z)\right)^2.
\label{eq:slack_ladder}
\end{equation}
When $U_s(x)>R_s$, the surplus ladder can absorb the excess. When $U_s(x)<R_s$, nonnegative surplus slack cannot eliminate the shortfall, so a penalty remains. This creates a compact quantum-compatible approximation to a one-sided coverage constraint.

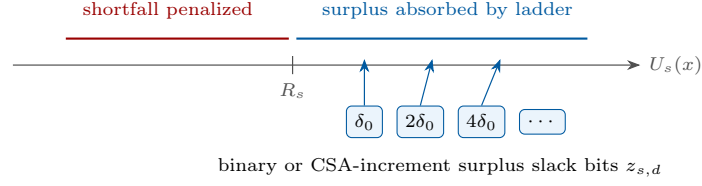
\begin{figure}[t]
\centering
\begin{tikzpicture}[
  axis/.style={-Stealth, line width=0.45pt, pygray},
  tick/.style={line width=0.45pt, pygray},
  lab/.style={font=\scriptsize},
  box/.style={draw=pyblue, fill=pylight, rounded corners=2pt, align=center, font=\scriptsize, inner sep=3pt}
]
  \draw[axis] (-0.2,0) -- (8.1,0) node[right, lab] {$U_s(x)$};
  \draw[tick] (3.5,0.12) -- (3.5,-0.12) node[below, lab] {$R_s$};
  \draw[draw=red!60!black, line width=0.8pt] (0.5,0.35) -- (3.45,0.35);
  \node[lab, red!60!black] at (1.85,0.72) {shortfall penalized};
  \draw[draw=pyblue, line width=0.8pt] (3.55,0.35) -- (7.4,0.35);
  \node[lab, pyblue] at (5.55,0.72) {surplus absorbed by ladder};
  \node[box] (s0) at (4.45,-0.75) {$\delta_0$};
  \node[box, right=0.18cm of s0] (s1) {$2\delta_0$};
  \node[box, right=0.18cm of s1] (s2) {$4\delta_0$};
  \node[box, right=0.18cm of s2] (s3) {$\cdots$};
  \draw[-Stealth, pyblue, line width=0.45pt] (s0.north) -- (4.45,0.08);
  \draw[-Stealth, pyblue, line width=0.45pt] (s1.north) -- (5.35,0.08);
  \draw[-Stealth, pyblue, line width=0.45pt] (s2.north) -- (6.25,0.08);
  \node[lab] at (5.45,-1.35) {binary or CSA-increment surplus slack bits $z_{s,d}$};
\end{tikzpicture}
\caption{One-sided coverage slack ladder. The model penalizes margin shortfall while allowing discrete surplus to be represented by nonnegative slack bits.}
\label{fig:slack_ladder}
\end{figure}

The denominations can be binary powers, CSA rounding increments, or collateral lot increments:
\begin{equation}
\delta_d=2^d\delta_0,\qquad d=0,\ldots,D_s.
\end{equation}
Figure~\ref{fig:slack_ladder} illustrates the one-sided encoding. Surplus coverage can be represented by slack denominations; true shortfall remains exposed to the penalty.

The slack ladder is particularly important for SIMM IM because IM collateral is often segregated and harder to reuse; surplus IM is economically more costly than surplus VM, but the model still distinguishes surplus from true shortfall.

\section{Higher-Order QAOA Ansatz}
\label{sec:qaoa}

\subsection{Phase Separator}
Given the Hamiltonian in Eq.~\eqref{eq:cost_hamiltonian}, the CR-HO-QAOA phase separator is
\begin{equation}
U_C(\gamma)=\exp(-i\gamma \hat{H}_C).
\end{equation}
Because all Pauli-$Z$ strings commute, the phase separator can be decomposed into a product of string phases:
\begin{equation}
U_C(\gamma)=\prod_{e\in\mathcal{E}}\prod_{S\subseteq e}
\exp\left(-i\gamma\theta_{e,S}\prod_{j\in S}Z_j\right),
\end{equation}
where
\begin{equation}
\theta_{e,S}=\frac{c_e}{2^{|e|}}(-1)^{|S|}.
\end{equation}
For a Pauli string $Z_{j_1}\cdots Z_{j_m}$, the unitary
\begin{equation}
\exp(-i\alpha Z_{j_1}\cdots Z_{j_m})
\end{equation}
can be implemented with a parity network: compute parity onto one qubit using CNOT gates, apply a single $R_Z(2\alpha)$ rotation, and uncompute. For string length $m>1$, this requires $2(m-1)$ CNOT gates and one $R_Z$ rotation before hardware transpilation. For $K\le4$, the highest-order native string is at most four-local.

For example, a four-local phase $\exp(-i\alpha Z_aZ_bZ_cZ_d)$ on a linear qubit order $(a,b,c,d)$ can be compiled as the following parity network:
\begin{figure}[t]
\centering
\fbox{\begin{minipage}{0.88\linewidth}
\textbf{Four-local parity phase decomposition.}
\begin{enumerate}
    \item Apply $\mathrm{CNOT}(a,b)$, $\mathrm{CNOT}(b,c)$, $\mathrm{CNOT}(c,d)$ to accumulate parity on qubit $d$.
    \item Apply $R_Z(2\alpha)$ on qubit $d$.
    \item Apply $\mathrm{CNOT}(c,d)$, $\mathrm{CNOT}(b,c)$, $\mathrm{CNOT}(a,b)$ to uncompute the parity.
\end{enumerate}
This implementation uses six CNOT gates and one $R_Z$ rotation before routing overhead. On a statevector simulator the same operation can be applied directly as a diagonal phase; on hardware the sequence is transpiled to the device coupling map.
\end{minipage}}
\caption{Concrete decomposition of a four-local Pauli-$Z$ phase used by the higher-order phase separator.}
\label{fig:four_local_decomp}
\end{figure}

When a target backend does not support efficient high-order phase synthesis, the same HOBO instance may be quadratized into a QUBO using auxiliary variables. However, quadratization is treated as a backend compilation step, not as the primary modeling layer. This preserves the higher-order collateral semantics during formulation and analysis.

\subsection{Mixer Families}
The framework proposes collateral-specific structured-subspace mixer families. These are formulation designs rather than claims about the executed confirmatory mixer. In the confirmatory experiment, CR-HO-QAOA uses the original builder idle-XY mixer and is compared against a matched builder X-mixer on the same Hamiltonian. Gate~B1 passes for the registered encoded at-most-one property, whereas Gate~B2 fails for full joint collateral-constraint preservation. The proposed mixer family sum is:
\begin{equation}
\hat{H}_M=\hat{H}_{onehot}+\hat{H}_{card}+\hat{H}_{side}+\hat{H}_{sub}+\hat{H}_{batch}.
\label{eq:mixer_sum}
\end{equation}
The corresponding mixer unitary is
\begin{equation}
U_M(\beta)=\exp(-i\beta \hat{H}_M).
\end{equation}

\paragraph{One-hot eligibility mixer.}
Suppose a group $G$ contains mutually exclusive alternatives, such as block-size choices for one asset or replacement choices for one recalled position. Feasibility requires
\begin{equation}
\sum_{i\in G}x_i=1.
\end{equation}
An XY mixer over the group preserves this Hamming weight:
\begin{equation}
\hat{H}_{XY}^{G}=\sum_{(i,j)\in E_G}\frac{1}{2}(X_iX_j+Y_iY_j),
\label{eq:xy_mixer}
\end{equation}
where $E_G$ may be a ring, line, or clique topology. A clique gives richer mixing but higher depth; a ring is shallower and more hardware-friendly.

\paragraph{Cardinality movement-budget mixer.}
Collateral desks often impose a maximum or target number of operational movements. For a movement group $B$, a fixed-cardinality mixer preserves
\begin{equation}
\sum_{i\in B}x_i=K_B.
\end{equation}
The same XY construction in Eq.~\eqref{eq:xy_mixer} can be used with a Dicke-state or hot-started initialization inside the fixed-Hamming-weight subspace.

\paragraph{Side-preserving VM/IM mixer.}
VM and IM collateral usually have different legal and economic treatment. Let $\mathcal{V}_{VM}$ and $\mathcal{V}_{IM}$ be side-specific action sets. The side mixer is block diagonal:
\begin{equation}
\hat{H}_{side}=\hat{H}_{XY}^{\mathcal{V}_{VM}}\oplus \hat{H}_{XY}^{\mathcal{V}_{IM}},
\end{equation}
so that VM actions do not mix into IM actions unless the legal state explicitly permits cross-side substitution.

\paragraph{Substitution-pair mixer.}
A collateral substitution often has a paired structure: recall an expensive posted block and post a cheaper eligible replacement. For substitution pairs $(r,p)$, define local alternatives that preserve the net substitution logic. The mixer connects alternatives only when the resulting recall-post pair remains inventory- and side-compatible. This avoids sampling states that recall collateral without replacing it or post replacement collateral without releasing the corresponding expensive block.

\paragraph{Batch mixer.}
Custody and triparty workflows create batch effects. A batch variable $b_m$ is activated when one or more actions in batch $m$ are selected. The batch mixer toggles feasible alternatives within a batch while respecting the logical activation relation. In near-term implementations, the batch logic may be enforced in the phase Hamiltonian while the mixer operates on the action variables.

\subsection{Warm-Started Initial States}
Let $x^0$ be the binary encoding of the current or CP-SAT feasible allocation. For unconstrained variables, a biased product warm start is
\begin{equation}
|\psi_0\rangle=\bigotimes_i \left(\sqrt{1-p_i}|0\rangle+\sqrt{p_i}|1\rangle\right),
\end{equation}
where $p_i$ is derived from action desirability, reduced cost, or similarity to $x_i^0$. For one-hot groups, the initial state is concentrated on the incumbent alternative and its nearest economically plausible substitutions:
\begin{equation}
|\psi_0^G\rangle=\sum_{i\in G}\alpha_i|e_i\rangle,
\qquad \sum_{i\in G}|\alpha_i|^2=1.
\end{equation}
For fixed-cardinality movement budgets, a Dicke state or approximate Dicke state may be used. The practical goal is not to prepare a perfectly uniform feasible state, but to concentrate amplitude on business-relevant reallocations.

\subsection{Variational State and Optimization Objective}
At depth $p$, the CR-HO-QAOA state is
\begin{equation}
|\psi_p(\boldsymbol{\gamma},\boldsymbol{\beta})\rangle
=
\prod_{\ell=1}^{p}U_M(\beta_\ell)U_C(\gamma_\ell)|\psi_0\rangle.
\end{equation}
The standard objective is energy minimization:
\begin{equation}
\min_{\boldsymbol{\gamma},\boldsymbol{\beta}}
\langle\psi_p|\hat{H}_C|\psi_p\rangle.
\end{equation}
For collateral, a CVaR-style objective over the best sampled tail can be more robust:
\begin{equation}
\min_{\boldsymbol{\gamma},\boldsymbol{\beta}}\mathrm{CVaR}_{\alpha}\left(C(x)\right),
\end{equation}
where $\alpha$ is the retained quantile of low-cost samples. A multi-objective variant sweeps treasury weights
\begin{equation}
\omega=(w_F,w_M,w_L,w_O,w_I,w_C,w_S,w_G)
\end{equation}
across funding, movement and substitution operations, liquidity, overshoot, IM segregation and non-reuse, concentration, shortfall diagnostics, and governance costs. Positive prohibited shortfall remains disqualifying under certification rather than being traded against the other components. This weight sweep is a design extension for Pareto reporting and is not a claim of multi-objective superiority in the confirmatory package.

\section{CR-HO-QAOA Algorithm}
\label{sec:algorithm}

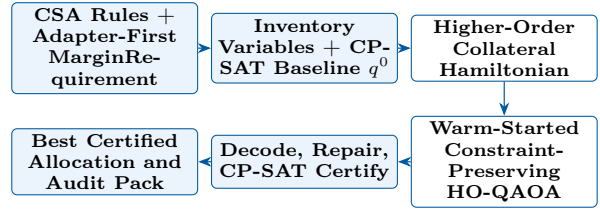
\begin{figure}[t]
\centering
\begin{tikzpicture}[
  node distance=0.17cm,
  box/.style={draw=pyblue, fill=pylight, rounded corners=2pt,
    minimum height=0.70cm, text width=0.19\linewidth, align=center,
    font=\scriptsize\bfseries, inner sep=2pt},
  qbox/.style={draw=pyblue, fill=white, rounded corners=2pt,
    minimum height=0.70cm, text width=0.19\linewidth, align=center,
    font=\scriptsize\bfseries, inner sep=2pt},
  arrow/.style={-Stealth, line width=0.45pt, pyblue}
]
  \node[box] (route) {CSA Rules + Adapter-First MarginRequirement};
  \node[box, right=of route] (base) {Inventory Variables + CP-SAT Baseline $q^0$};
  \node[qbox, right=of base] (hubo) {Higher-Order Collateral Hamiltonian};
  \node[qbox, below=0.45cm of hubo] (qaoa) {Warm-Started Structured-Subspace HO-QAOA};
  \node[box, left=of qaoa] (cert) {Decode, Repair, Deterministically Certify};
  \node[box, left=of cert] (out) {Best Certified Allocation and Audit Pack};
  \draw[arrow] (route) -- (base);
  \draw[arrow] (base) -- (hubo);
  \draw[arrow] (hubo) -- (qaoa);
  \draw[arrow] (qaoa) -- (cert);
  \draw[arrow] (cert) -- (out);
\end{tikzpicture}
\caption{CR-HO-QAOA is a quantum local-search layer embedded after adapter-first margin normalization and before deterministic certification. SIMM or legacy margin calculation is upstream of the optimizer.}
\label{fig:algorithm_flow}
\end{figure}

Figure~\ref{fig:algorithm_flow} summarizes the algorithmic pipeline. The adapter-first margin layer creates a normalized requirement from SIMM, external official SIMM, proxy, or legacy inputs; CSA terms and inventory create collateral variables; the quantum layer performs higher-order local search; and deterministic certification gates the outputs.

\subsection{Algorithm Steps: Candidate Decoding and Certification}

\begin{figure}[t]
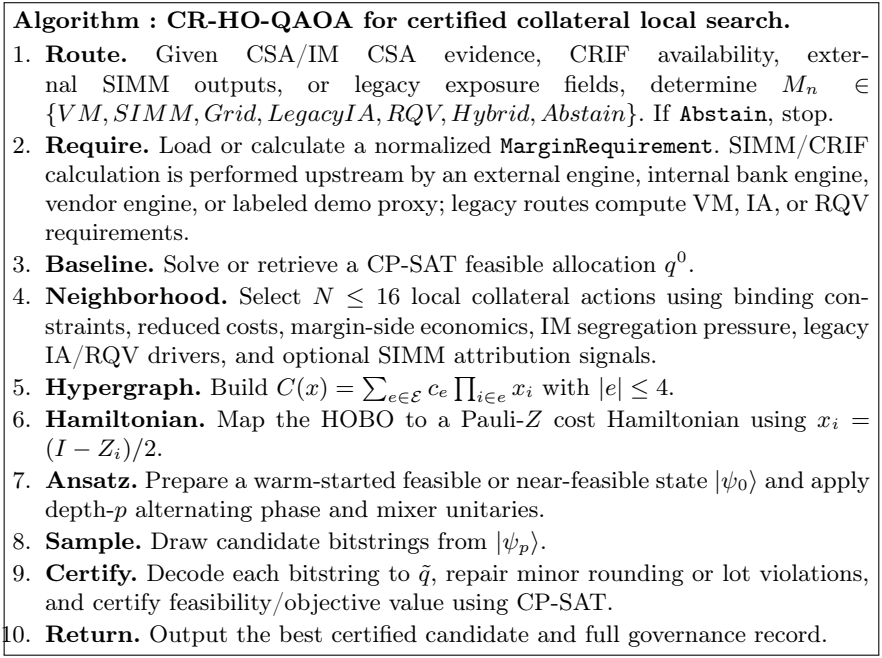

\centering
\fbox{\begin{minipage}{0.93\linewidth}
\textbf{Algorithm : CR-HO-QAOA for certified collateral local search.}
\begin{enumerate}
    \item \textbf{Route.} Given CSA/IM CSA evidence, CRIF availability, external SIMM outputs, or legacy exposure fields, determine $M_n\in\{VM,SIMM,Grid,LegacyIA,RQV,Hybrid,Abstain\}$. If \texttt{Abstain}, stop.
    \item \textbf{Require.} Load or calculate a normalized \texttt{MarginRequirement}. SIMM/CRIF calculation is performed upstream by an external engine, internal bank engine, vendor engine, or labeled demo proxy; legacy routes compute VM, IA, or RQV requirements.
    \item \textbf{Baseline.} Solve or retrieve a CP-SAT feasible allocation $q^0$.
    \item \textbf{Neighborhood.} Select $N\le16$ local collateral actions using binding constraints, reduced costs, margin-side economics, IM segregation pressure, legacy IA/RQV drivers, and optional SIMM attribution signals.
    \item \textbf{Hypergraph.} Build $C(x)=\sum_{e\in\mathcal{E}}c_e\prod_{i\in e}x_i$ with $|e|\le4$.
    \item \textbf{Hamiltonian.} Map the HOBO to a Pauli-$Z$ cost Hamiltonian using $x_i=(I-Z_i)/2$.
    \item \textbf{Ansatz.} Prepare a warm-started feasible or near-feasible state $|\psi_0\rangle$ and apply depth-$p$ alternating phase and mixer unitaries.
    \item \textbf{Sample.} Draw candidate bitstrings from $|\psi_p\rangle$.
    \item \textbf{Certify.} Decode each bitstring to $\tilde{q}$, repair only documented deterministic cases, and apply the implemented deterministic constraint and objective checks. A production deployment submits accepted candidates to its approved CP-SAT master.
    \item \textbf{Return.} Output the best certified candidate and full governance record.
\end{enumerate}
\end{minipage}}
\caption{CR-HO-QAOA algorithm.}
\label{fig:algo}
\end{figure}

A sampled bitstring $x$ produces a candidate allocation
\begin{equation}
\tilde{q}=q^0+\Delta q(x).
\end{equation}
The candidate is first checked against local logical constraints. A deterministic repair step may adjust lot-size, rounding, or surplus-slack consistency only if the repair is uniquely defined and economically immaterial. Let $\mathcal{F}_{\mathrm{impl}}$ denote the constraint set actually implemented by the certifier. The accepted candidate set is
\begin{equation}
\mathcal{C}_{acc}=\{\tilde{q}:\tilde{q}\in\mathcal{F}_{\mathrm{impl}}
\text{ and passes the audit checks}\}.
\end{equation}
In production, this gate is followed by the approved CP-SAT master. In the confirmatory package, all enumerable action states are pre-evaluated using the deterministic checker; samples use cached lookup, and the resulting decisions are cross-validated against fixed-allocation CP-SAT for the implemented overlapping constraints.
The production output is
\begin{equation}
q^{\star}=\arg\min_{q\in\mathcal{C}_{acc}}J(q).
\end{equation}
If $\mathcal{C}_{acc}$ is empty, the system returns the original CP-SAT baseline $q^0$ with a warning that the quantum local search did not improve the certified allocation.

\subsection{Algorithm Results}
The main algorithmic upgrade is summarized earlier in Table~\ref{tab:old_new}. CR-HO-QAOA changes the modeling primitive: the local collateral neighborhood is represented as a higher-order hypergraph and the phase separator acts on the induced Pauli-$Z$ strings. The proposed mixer families encode selected structural subspaces, but the executed builder establishes only the registered at-most-one preservation property and fails the joint-preservation gate. Section~\ref{sec:benchmark} therefore compares the original builder with matched X-mixer QAOA and native HOBO local search under the same frozen certification and candidate budget, without assuming that ansatz structure improves performance.

\section{SIMM-Aware Quantum Objective Design}
\label{sec:simm_quantum}

SIMM enters the quantum optimization problem through requirements, costs, eligibility, and neighborhood selection. It does not enter by allowing the optimizer to alter SIMM sensitivities or margin methodology. A CRIF row may be abstractly represented as
\begin{equation}
g_{j,r,\tau,b,q},
\end{equation}
where $j$ indexes the trade, $r$ the SIMM risk class, $\tau$ the risk type, $b$ the bucket, and $q$ the qualifier. The approved margin engine or external SIMM adapter maps the CRIF set $\mathcal{S}_n$ to $R_n^{IM}$. The quantum optimizer then minimizes the cost of satisfying $R_n^{IM}$ together with any VM or legacy IA requirement.

The design deliberately supports two deployment modes. In bank integration mode, $R_n^{IM}$ is supplied by an official internal or vendor SIMM engine and is accompanied by metadata such as SIMM version, valuation date, CRIF hash, source system, and approval status. In research mode, a CRIF-driven demo proxy may supply a non-official IM proxy for validating the optimization pipeline. The latter must be labeled as a proxy result and is not used as evidence of official SIMM calculation. In both modes, the downstream quantum and CP-SAT layers consume only the normalized requirement and source metadata.

The IM funding component can be written as
\begin{equation}
C_F^{IM}(x)=\sum_i a_{i,IM}\Delta q_{i,IM}(x)
\left(r_i+\ell_i+s_{xccy,i}+s_{seg,i}+s_{nonreuse,i}\right)
\end{equation}
where $s_{seg,i}$ captures segregation drag and $s_{nonreuse,i}$ captures the opportunity cost of collateral that cannot be reused. This term is side-specific: the same collateral action can have different economics under VM and IM.

A SIMM-aware neighborhood score for action $i$ may be defined as

\begin{equation}
\begin{aligned}
\sigma_i ={}&
\alpha_1\mathrm{ReducedCost}_i
+\alpha_2\mathrm{IMCost}_i
+\alpha_3\mathrm{SegDrag}_i \\
&+\alpha_4\mathrm{BucketRelevance}_i
+\alpha_5\mathrm{CapPressure}_i 
\end{aligned}
\label{eq:action_score}
\end{equation}

where $\mathrm{BucketRelevance}_i$ uses SIMM attribution or desk-level IM decomposition only to prioritize local collateral actions. It does not modify the SIMM calculation itself.

This design gives the quantum layer a meaningful SIMM-specific role: it explores the discrete collateral reallocation space created by IM requirements, segregation economics, and side-specific eligibility. The regulatory margin model remains outside the variational loop.

\section{Circuit Resource and Execution Considerations}
\label{sec:resources}

The proposed quantum layer is intentionally local. For $N\le16$ active actions and maximum hyperedge order $K\le4$, the model remains accessible to exact or reduced-sector statevector simulation and quantum-inspired analysis. Executability on a physical processor is not established by this work. The relevant resource drivers are:
\begin{itemize}
    \item number of action qubits $N$ plus idle/reservoir and other auxiliary qubits;
    \item number of hyperedges $|\mathcal{E}|$;
    \item maximum hyperedge order $K$;
    \item Pauli-string count after expansion;
    \item mixer topology depth;
    \item QAOA depth $p$;
    \item shot budget and optimizer iterations.
\end{itemize}

For a hyperedge of order $m$, Eq.~\eqref{eq:pauli_expansion} produces up to $2^m$ Pauli strings, though constant terms and duplicate strings can be compressed. A Pauli string of length $m$ can be implemented with $2(m-1)$ CNOT gates and one $R_Z$ rotation using a parity-network decomposition. Thus, high-order terms are manageable only when the active neighborhood and hyperedge count are bounded.

The frozen confirmatory performance protocol uses shallow depths $p\in\{1,2\}$. A future scaling protocol may study
\begin{equation}
p\in\{1,2,3\},
\end{equation}
with parameter optimization by COBYLA, SPSA, grid transfer, or learned schedules from smaller instances. Any hardware demonstration must separately report transpiled two-qubit gate counts, complete state preparation, circuit depth, shot budget, noise model, and backend topology.

The intended claim is not near-term quantum advantage over CP-SAT. The intended claim of this manuscript is a reproducible higher-order formulation and certification architecture. The confirmatory package shows that original builder idle-XY does \emph{not} beat matched X-mixer QAOA on the registered primary certified metric, and that native HOBO local search remains stronger on the registered classical comparison.

\section{Confirmatory Benchmark Protocol and Results}
\label{sec:benchmark}

This section replaces earlier exploratory synthetic ablation tables with the archived confirmatory record. The repository is
\url{https://github.com/Pyligent/collateral_hybrid}.
The performance package is archived under
\path{results/cr_improvement_v6} at tag \texttt{cr-v6-freeze}.
The PennyLane cross-validation package is archived under
\path{results/cr_improvement_v6_2_validation} at tags
\texttt{cr-v6-2-xv-freeze} and \texttt{cr-v6-2-xv-results}.
Earlier exploratory packages (v1--v5) are development history only and are not used as confirmatory performance evidence.

\begin{table}[t]
\centering
\caption{Confirmatory protocol (archived v6 performance package).}
\label{tab:protocol}
\scriptsize
\begin{tabular}{@{}ll@{}}
\toprule
Item & Value \\
\midrule
Protocol id & \texttt{cr\_ho\_qaoa\_v6} \\
Primary metric & \texttt{final\_certified\_improving\_probability} \\
Primary neighborhood & $N_{\mathrm{requested}}=N_{\mathrm{eff}}=8$ \\
Depth grid & $p\in\{1,2\}$ \\
Seeds / fixture-config & 30 \\
Seed-level rows & 6300 \\
Primary candidate budget & 16 \\
Primary confirmatory fixtures & 11 (six families; $N_{\mathrm{eff}}=8$) \\
Exact local ground truth & full neighborhood enumeration \\
Sample certification & cached deterministic enumeration table \\
CP-SAT role & reference solve and fixed-allocation cross-check \\
Matched quantum control & builder X-mixer QAOA \\
Native classical baseline & HOBO local search \\
Simulator & noise-free NumPy statevector; shots $=512$ \\
Software & Python 3.12.12; NumPy 2.5.1; PennyLane 0.45.1 \\
Freeze commit & \texttt{09a0a6dd\ldots} / tag \texttt{cr-v6-freeze} \\
\bottomrule
\end{tabular}
\end{table}

\subsection{Primary metric and controls}
The primary metric is \texttt{final\_certified\_improving\_probability}: the exact probability mass, under the final simulated distribution, on action states that pass the implemented deterministic certification mask and improve the production objective relative to the incumbent. All $2^8=256$ action states per fixture are decoded, evaluated, and pre-certified before sampling metrics are calculated. Per-sample scoring uses the resulting frozen lookup table; CP-SAT is not invoked once per sampled candidate. Fixed-allocation CP-SAT is used for the reference solve and to cross-check the implemented overlapping feasibility decisions, producing 256/256 agreement on every primary fixture. Exact neighborhood enumeration supplies local ground truth. Matched builder X-mixer QAOA is the quantum control. Native HOBO local search is the classical higher-order baseline. Performance methods use $p\in\{1,2\}$, CVaR variational training with $\alpha=0.2$, optimizer budget 40 iterations, and 512 shots for sampling metrics. Funding-cost changes use
\begin{equation}
\Delta C_F =
C_F(\text{certified allocation})
-
C_F(\text{current allocation}),
\end{equation}
so negative $\Delta C_F$ means lower funding cost. Positive changes for infeasible baselines are feasibility-restoration costs, not ``negative savings.'' Weighted-objective reduction is not cash funding savings. The confirmatory packages do not contain a provenance-complete funding-savings table; numerical production-savings claims are therefore omitted.

\subsection{Implemented certification coverage}
Table~\ref{tab:constraint_coverage} distinguishes hard checks from objective, decoding, and metadata treatment. The reported CP-SAT equivalence result applies to the overlapping implemented constraint set; it is not evidence that every proposed production rule is already encoded as a hard predicate.

\begin{table}[t]
\centering
\caption{Constraint coverage in the frozen confirmatory implementation.}
\label{tab:constraint_coverage}
\scriptsize
\begin{tabular}{@{}p{0.25\linewidth}p{0.31\linewidth}p{0.32\linewidth}@{}}
\toprule
Control & Confirmatory handling & Fixed-allocation CP-SAT relation \\
\midrule
VM/IM coverage, inventory, side eligibility, issuer concentration
& Deterministic hard check
& Explicitly checked; exact decision agreement \\
Movement budget
& Hard action-bit Hamming bound
& Not a master-model constraint \\
Lot size and rounding
& Integer decode/valuation path
& Integer quantities in fixed solve \\
Zero shortfall and $P_G=0$
& Deterministic rejection
& Aligned post hoc for equivalence test \\
Substitution and batch logic
& Economics in HOBO/objective; no dedicated hard predicate
& Not implemented as a hard constraint \\
Custody
& Certificate metadata only
& Not feasibility-checked \\
Segregation and reuse
& $P_I$ objective component
& Not implemented as a hard constraint \\
\bottomrule
\end{tabular}
\end{table}

\subsection{Decision gates}
\begin{table}[t]
\centering
\caption{Preregistered decision gates for the confirmatory package. Pass/fail values are taken from \texttt{results/cr\_improvement\_v6/aggregate/decision\_gates.json}.}
\label{tab:gates}
\scriptsize
\begin{tabular}{@{}clp{0.28\textwidth}p{0.38\textwidth}@{}}
\toprule
Gate & Outcome & Definition & Evidentiary role \\
\midrule
A & FAIL & Builder Hamiltonian alignment on registered supports & Structural prerequisite; does not support builder optimization claims \\
B1 & PASS & Encoded at-most-one preservation by executed mixer & Supports unitary preservation of the registered AMO structure only \\
B2 & FAIL & Full claimed joint collateral-constraint preservation & Does not support unconstrained constraint-preserving wording \\
C & PASS & Release hygiene / experiment-source integrity & Supports reproducibility packaging, not performance \\
D1 & FAIL & Builder idle-XY vs matched X on primary certified metric & Primary negative result: X favored on 9/11 fixtures \\
D2 & PASS & Secondary mechanism / difference-in-differences check & Secondary only; must not overturn D1/A/F failures \\
D3 & FAIL & Improving-sample rate at candidate budget 16 & Idle-XY mean rate below matched X \\
E & PASS & M\"obius $K{=}4$ vs $K{=}2$ diagnostic idle-XY & Diagnostic only; not a builder higher-order advantage \\
F & FAIL & Builder idle-XY vs native HOBO local search & Native HOBO remains stronger on registered comparison \\
G & PASS & Builder vs M\"obius diagnostic reporting gate & Reporting/diagnostic; M\"obius not a scalable builder \\
\bottomrule
\end{tabular}
\end{table}

\subsection{Builder idle-XY versus matched X and native HOBO}
\begin{table}[t]
\centering
\caption{Builder idle-XY versus matched X-mixer and native HOBO local search on the registered primary confirmatory fixtures ($N_{\mathrm{eff}}=8$). Aggregates from \texttt{decision\_gates.json} Gates D1, D3, and F.}
\label{tab:methods}
\scriptsize
\begin{tabular}{@{}lrr@{}}
\toprule
Comparison & Statistic & Value \\
\midrule
Idle-XY vs matched X (Gate D1) & Fixtures favoring idle-XY & 2 \\
 & Fixtures favoring X & 9 \\
 & Mean fixture effect (idle$-$X) & $-0.04857$ \\
 & Equivalent percentage points & $-4.86$ \\
Improving-sample rate at budget 16 (Gate D3) & Mean idle-XY (\%) & $12.12$ \\
 & Mean matched X (\%) & $16.85$ \\
Idle-XY vs native HOBO LS (Gate F) & Fixtures favoring idle-XY & 1 \\
 & Fixtures favoring HOBO LS & 10 \\
 & Mean rate difference (idle$-$HOBO) & $-10.247$ \\
\bottomrule
\end{tabular}
\end{table}

\paragraph{Primary negative result.}
Gate~A fails: the original builder Hamiltonian fails alignment on the registered supports. Gate~D1 fails: builder idle-XY favors only 2 of 11 fixtures while matched X favors 9 of 11; the mean fixture effect is $-0.0485667501781553$ ($\approx -4.86$ percentage points). Gate~D3 fails: at candidate budget 16, mean improving-sample rates are $12.120028409090908\%$ (idle-XY) versus $16.853693181818183\%$ (matched X). Gate~F fails: native HOBO local search remains stronger (HOBO favored on 10/11 fixtures).

\paragraph{Secondary and diagnostic gates.}
Gate~B1 passes and Gate~B2 fails, so unitary constraint claims are limited to the registered at-most-one structure. Gate~D2 passes only as a secondary mechanism check and must not be presented as a primary performance win. Gate~E passes for diagnostic M\"obius $K{=}4$ versus $K{=}2$ idle-XY variants; M\"obius Hamiltonians are enumeration-based diagnostics, not scalable replacements for the original builder, and Gate~E must not be used to claim a general higher-order advantage. Gate~G reports builder-versus-M\"obius diagnostics without overturning A/D1/D3/F. Gate~C passes for experiment-source release hygiene.

\subsection{PennyLane cross-validation}
\begin{table}[t]
\centering
\caption{PennyLane \texttt{default.qubit} cross-validation summary (archived v6.2 validation package; tag \texttt{cr-v6-2-xv-results}). Fixed $p=1$ parameters $(\gamma,\beta)=(0.42,0.31)$.}
\label{tab:xv}
\scriptsize
\begin{tabular}{@{}lr@{}}
\toprule
Item & Value \\
\midrule
Fixtures / rows & 11 / 11 \\
Release status & \texttt{READY\_FOR\_INDEPENDENT\_REVIEW} \\
Minimum fidelity & $0.9999999999999991$ \\
Max amplitude difference & $0.0$ \\
Max probability difference & $0.0$ \\
Max expectation difference & $0.0$ \\
Max action-marginal $\ell_1$ & $0.0$ \\
Max normalization residual & $4.441e-16$ \\
Total qubits & 14--15 \\
NumPy mode & reduced-sector evolution, embedded to full Hilbert space \\
\bottomrule
\end{tabular}
\end{table}

For eleven registered $N_{\mathrm{eff}}=8$ fixtures, the custom NumPy reduced-sector simulator was cross-validated against PennyLane \texttt{default.qubit} at fixed $p=1$ parameters $(\gamma,\beta)=(0.42,0.31)$. The reduced-sector NumPy states were embedded into the full 14--15-qubit Hilbert space before comparison. All eleven comparisons passed the preregistered tolerances: minimum fidelity was $0.9999999999999991$, while maximum amplitude, probability, expectation, and action-marginal differences were zero to the reported precision. Normalization residuals were at numerical roundoff scale.

This validates the implemented mathematical state evolution for the registered circuits. It is not hardware validation, noisy simulation, transpilation validation, or validation of a complete state-preparation and gate-decomposition pipeline. NumPy execution is reduced-sector evolution embedded into the full state for comparison, not ``full-mode simulation.''

\subsection{Circuit-resource scope}
\begin{table}[t]
\centering
\caption{Circuit-resource scope for confirmatory $N_{\mathrm{eff}}=8$ builder methods. Counts are analytical QAOA evolution-only estimates; structured state preparation and hardware transpilation are excluded.}
\label{tab:resources}
\scriptsize
\begin{tabular}{@{}lrrrrr@{}}
\toprule
Method & $p$ & Action & Idle/aux & Total & 2Q-equivalent estimate \\
\midrule
Builder idle-XY (typical) & 1 & 8 & 6 & 14 & 422 \\
Builder idle-XY (HO20 VM) & 1 & 8 & 7 & 15 & 442 \\
Matched builder X & 1 & 8 & 0 & 8 & 402 \\
\bottomrule
\end{tabular}
\end{table}

These values are not measured compiled-circuit depths or physical gate counts. Pauli terms are scheduled from stored term orders using artificial round-robin qubit supports because exact Pauli supports are not retained by the reporting path. Abstract graph-mixer transitions also lack a complete qubit-gate decomposition. Structured state preparation, routing, hardware transpilation, and noise are excluded; the table is therefore suitable only for relative analytical bookkeeping.

\section{Discussion}
\label{sec:discussion}

The proposed framework remains a quantum-finance application for three reasons. First, the core optimization object is a cost Hamiltonian over collateral-action qubits. Second, higher-order collateral interactions are represented as Pauli-$Z$ phase terms. Third, the mixer design is informed by structured-subspace quantum dynamics rather than by classical penalty-only modeling alone. The financial domain determines Hamiltonian terms, warm starts, and certification constraints through the adapter-first margin boundary.

Matched X can outperform structured idle-XY after certification for several reinforcing reasons. Strict subspace restriction reduces the reachable distribution relative to a transverse-field mixer. The builder Hamiltonian is poorly aligned with the certified production objective (Gate~A), so concentrating amplitude inside a structured sector need not improve certified improving probability. Preservation of one structural subset (Gate~B1) does not guarantee better economic candidates when joint constraints fail Gate~B2. The certification layer can rescue candidates generated outside an intended mixer sector, decoupling ansatz structure from certified value. Ansatz structure is therefore not itself evidence of optimization value.

Diagnostic M\"obius variants that improve Hamiltonian alignment on enumerable instances are useful for analysis but are not presented as production builders. The confirmatory contribution is therefore the formulation, simulator implementation, and governed certification architecture---not an optimization advantage for the original CR-HO-QAOA builder.

\section{Limitations and Future Work}
\label{sec:limitations}

The confirmatory study uses synthetic fixtures and classically enumerable neighborhoods; there is no quantum-hardware execution and no noise model. Depth is restricted to small fixed $p\in\{1,2\}$ for performance and $p=1$ for PennyLane validation. Only eleven primary confirmatory fixtures at $N_{\mathrm{eff}}=8$ support the idle-versus-X claim; the proposed $N\in\{8,12,16\}$ design grid is not a completed confirmatory experiment. Circuit-resource totals exclude structured state preparation and backend transpilation, use artificial round-robin supports, and are not measured compiled-circuit counts. PennyLane validation is functional statevector validation of the registered mathematical evolution. Per-sample certification uses a cached deterministic table, not a fresh CP-SAT invocation, and the equivalence result covers only overlapping implemented constraints. Substitution and batch consistency are not dedicated hard predicates, custody is metadata-only, and segregation/non-reuse is represented in the objective. The builder Hamiltonian fails Gate~A alignment; Gate~B2 fails; there is no advantage over matched X or native HOBO local search; there is no clean higher-order advantage for the original builder; and there is no production bank-savings evidence in the confirmatory packages.

This work also does not claim to implement or replace an approved ISDA SIMM engine. In production, the preferred architecture is adapter-first: Pyligent consumes official SIMM outputs from a bank, vendor, or approved internal engine and normalizes them into a margin requirement.

Future work has five directions. First, recalibrate or replace the builder Hamiltonian so that Gate~A alignment holds on mixer-reachable supports before further QAOA tuning. Second, enlarge neighborhoods beyond classical enumeration only after classical higher-order baselines remain in the protocol. Third, compare direct higher-order phase separators against quadratized QUBO circuits under equal hardware depth. Fourth, evaluate multi-objective reporting across funding, movement, and segregation costs without equating objective reduction to cash savings. Fifth, integrate with bank or vendor SIMM systems through an auditable external adapter and historical margin-call replay.

\section{Conclusion}
\label{sec:conclusion}

This manuscript presented an adapter-first, SIMM-aware collateral optimization framework in which the quantum component is a bounded local candidate generator and deterministic optimization remains the production authority. The main architectural takeaway is the separation of responsibilities: SIMM or legacy margin requirements are calculated or loaded upstream and normalized into a \texttt{MarginRequirement}; CSA terms define legal eligibility, haircuts, rounding, custody, and VM/IM side rules; inventory defines the available collateral universe; and CR-HO-QAOA searches a higher-order local action space whose candidates are decoded and evaluated before any recommendation. The confirmatory benchmark pre-certifies enumerable local states with the implemented deterministic checker and validates the overlapping decisions against fixed-allocation CP-SAT; a production deployment would submit recommended candidates to its approved master solver.

The study establishes a reproducible and exactly verifiable higher-order quantum formulation embedded within deterministic collateral certification. It does not establish an optimization advantage for the original CR-HO-QAOA builder or full implementation of every proposed collateral rule. Across the confirmatory fixtures, matched X-mixer QAOA and native classical higher-order search remain stronger on the registered metrics. The principal validated contribution is therefore the formulation, simulator implementation, explicit constraint-coverage audit, and governed certification architecture.


\end{document}